\documentclass[sigconf,10pt,nonacm]{acmart}
\usepackage{tikz}
\usepackage{multirow}
\usepackage{amsmath,amsfonts}

\usepackage{paralist}
\usepackage{url}

\usepackage{threeparttable}
\usepackage{booktabs}
\usepackage{makecell}
\usepackage{wasysym}
\usepackage{enumitem}
\usepackage{colortbl}
\usepackage{algorithm}
\usepackage[noend]{algpseudocode}
\usepackage{graphicx}
\usepackage{subcaption}
\usepackage{float}
\usepackage{xcolor}
\usepackage{xspace}

\def\BibTeX{{\rm B\kern-.05em{\sc i\kern-.025em b}\kern-.08em
    T\kern-.1667em\lower.7ex\hbox{E}\kern-.125emX}}

\newcommand{\mypara}[1]{\vspace{2pt}\noindent\textbf{#1 }}

\newtheorem{definition}{Definition}

\renewcommand{\AA}{\mathbf{A}}

\newcommand{\myPr}[1]{\ensuremath{\mathsf{Pr}\left[#1\right]}\xspace}

\title{Network-Aware Differential Privacy}

    \author{Zhou Li}
\affiliation{%
  \institution{University of California, Irvine}%
  \city{Irvine, CA}
  \country{USA}%
}
\author{Yu Zheng}
\affiliation{%
  \institution{University of California, Irvine}%
  \city{Irvine, CA}
  \country{USA}%
}  
\author{Tianhao Wang}
\affiliation{%
  \institution{University of Virginia}%
  \city{Charlottesville, VA}
  \country{USA}%
}  
\author{Sang-Woo Jun}
\affiliation{%
  \institution{University of California, Irvine}%
  \city{Irvine, CA}
  \country{USA}%
}  

\begin{abstract}
Differential privacy (DP) is a privacy-enhancement technology (PET) that receives prominent attention from the academia, industry, and government. One main development over the past decade has been the decentralization of DP, including local DP and shuffle DP.
Despite that decentralized DP heavily relies on network communications for data collection,
we found that: 1) no systematic study has surveyed the research opportunities at the intersection of networking and DP; 2) nor have there been significant efforts to develop DP mechanisms that are explicitly tailored for network environments.
In this paper, we seek to address this gap by initiating a new direction of network-aware DP. 
We identified two focus areas where the network research can offer substantive contributions to the design and deployment of DP, related to network security and topology. 
Through this work, we hope to encourage more research that  
adapt/optimize DP's deployment in various network environments. 
\end{abstract}

\begin{document}

\maketitle


\section{Introduction}
\label{sec:intro}

Differential privacy (DP)~\cite{dwork2006calibrating} has become one of the most prominent approaches to protect users' privacy under statistical data analysis and machine-learning applications in the recent decades. The involvement from academia and industry are strong and rapidly growing: more than thousands of papers about DP have been published since 2006~\cite{desfontainesblog20220309} and leading IT companies like Google and Apple have deployed DP into their running systems~\cite{apple-dp, erlingsson2014rappor}. 

In its earliest form, DP is mainly designed under a \textit{centralized database} setting, assuming the database server is trusted and the attacker aims to infer sensitive information by issuing queries. In the recent decade, there is a growing trend to \textit{decentralize} DP, with main variations like local DP and shuffle DP. In the decentralized setting, each client sends its reports (e.g., periodical measurement of sensor readings) to the central server, after a DP mechanism adds noises to them (e.g., randomized response~\cite{Warner65}). The central server then performs analytical tasks on the aggregated reports. By not trusting the central server, decentralized DP provides even better privacy protection. 

Through decentralizing DP and applying DP mechanisms in the data collection phase, now DP mechanisms also interleave with the communication networks.
From the perspective of network area, some questions would raise naturally: \textit{how are the network paradigms designed for decentralized DP?} or take a step back, \textit{are network protocols and environment considered when design DP mechanisms?}

Surprisingly, based on our literature review, we found the network factors were rarely considered or even mentioned by DP papers. 
We argue now it is a good time to work on the intersection between DP and network. Below we list a few motivations.

\begin{itemize}
    \item 
    Under the decentralized DP setting, DP mechanisms are executed by many distributed entities, rather than a single entity under central DP. As such, decentralized DP can be more vulnerable and the research of manipulation attack has gained prominent traction since its discovery in 2021~\cite{cheu2021manipulation, cao2021data}. These works all assume users are corrupted, but network adversaries are largely ignored, leaving a research gap unfilled.
    \item System-oriented works to optimize DP mechanisms, with a focus on the central DP setting~\cite{roth2020orchard, roth2021mycelium}, have seen prominent progress in the recent decade. Similarly, network-oriented optimization can be performed to improve the decentralized DP mechanisms.
    \item In the intersection of security and AI, we have seen numerous works on both directions: ``security for AI'' (e.g., research about data poisoning attacks) and ``AI for security'' (e.g., using unsupervised learning to detect intrusions). Regarding network and DP, we have seen some works about ``DP for network'', and the opposite direction ``network for DP'' should deserve more attention.
\end{itemize}

In this work, we discuss some \textbf{research opportunities (ROs)} that could be relevant under the theme of ``network for DP''. Our study follows a framework that identifies the techniques, innovations, and challenges developed or encountered by the network area, and draws connections with the existing DP mechanisms. 
So far, we have identified two focus areas that network research can contribute to, including examining DP under untrusted network and adapting DP to a network environment.
We found the techniques developed for network applications, like routing, caching, and service discovery, seem to be useful for DP mechanisms. 
Through this work, we hope to enlighten this under-studied direction, which we term as \textit{network-aware DP}.

\section{A Primer of Differential Privacy}
\label{sec:dp}

Traditional differential privacy (DP), also called central DP, assumes a trusted central server that collects raw reports from each client and adds noise to answer statistical queries (e.g., queries about the mean values)~\cite{dwork2006calibrating}. DP has a few important properties that propel its deployment. First, the privacy leakage under the malicious queries is \textit{provably bounded} by a privacy budget $\varepsilon$ under noisy mechanisms like Laplace mechanism. Second,  
DP is composable in the sense that applying multiple algorithms on a dataset under privacy budgets $\varepsilon_1, \cdots,\varepsilon_m$ satisfies $\varepsilon$-DP for $\varepsilon=\sum_{i} \varepsilon_i$.  
Third, after a DP mechanism is applied, it is always safe to perform arbitrary computations on the output because of DP's post-processing property.

Yet, the central server under standard DP might not always be trustworthy, and the server owner might want to infer sensitive information about an individual client. Another line of DP work focuses on the \textit{decentralized DP} setting to address this issue, which is also the focus of this paper. Below we elaborate two  main branches under decentralized DP.

\mypara{Local DP.}
Local DP asks clients to add noise directly to their reports before sending them to the central server, thus achieving a stronger privacy guarantee~\cite{erlingsson2014rappor}. The local DP notion ensures that any effect due to the contribution of one individual's data is limited (by the parameter $\varepsilon$).  More formally, 

\begin{definition}[{$\varepsilon$-Local Differential Privacy ($\varepsilon$-LDP)}] \label{def:ldp}
An algorithm $\AA$ satisfies $\varepsilon$-LDP if and only if for any pair of input values $v$ and $v'$ in the domain of $\AA$ and possible output $y \in \mathcal{Y}$, we have
\begin{equation}
\myPr{\AA(v) =y} \leq e^{\varepsilon}\, \myPr{\AA(v') =y}. \label{eq:ldp}
\end{equation}
where $\myPr{\cdot}$ denotes probability and $\varepsilon$ is privacy budget. A smaller $\varepsilon$ means stronger privacy protection.
\end{definition}

$\varepsilon$-LDP can be too strict sometimes and $(\varepsilon, \delta)$-LDP relaxes it by allowing $\AA$ to have a probability $\delta$ of not fulfilling $\varepsilon$-LDP.

The design of $\AA$ is key to the privacy-utility tradeoff (i.e., the resulting fidelity under a given privacy budget). Many algorithms have been designed for specific analytical tasks, such as frequency estimation and mean value estimation~\cite{Wang2017usenix}, heavy hitters discovery~\cite{tdsc:wang2019locally}, and federated learning~\cite{truex2020ldp}.
Local DP has gained prominent interests from the public and private sectors for privacy-preserving data collection and analysis, and the prominent systems that integrate Local DP include Apple iOS~\cite{apple-dp}, Google Chrome~\cite{erlingsson2014rappor}, Microsoft Windows system~\cite{ding2017collecting}, etc.
In addition, companies have also built libraries to help developers implement Local DP~\cite{diffprivlib,opacus,url:google-dp,url:smartnoise}.

\mypara{Shuffle DP.}
Though local DP removes the single point of trust (central server), it suffers prominent utility loss as a large amount of noise is added to each user's data. 
Shuffle DP~\cite{cheu2019distributed} was proposed to address the utility loss of local DP and achieve a middle ground between central DP and local DP. 
In this model, each user adds noise under a local DP mechanism, encrypts the data, and then sends it to a new intermediate server called \textit{shuffler}. The shuffler permutes the order of received reports and sends them to the central server. The central server decrypts the data and computes the aggregated statistics. In this process, the shuffler does not know the content of users' reports due to encryption, and the central server cannot link a report to the user due to permutation. 
Shuffle DP can achieve the {similar} error bounds as central DP in many applications: e.g., single-message shuffle DP achieves $\mathcal{O}(\frac{1}{\varepsilon})$ lower error
bound with $\mathcal{O}(\frac{\sqrt{n}}{\varepsilon})$ bits sent per user~\cite{cheu2021pure},  where $n$ is the number of users. Meanwhile, local DP would lead to the lower error
bound of $\mathcal{O}(\frac{\sqrt{n}}{\varepsilon})$ with the standard randomized response mechanism~\cite{Warner65}.

Like local DP, shuffle DP has also received signficant adoption by industry, such as Prochlo by Google~\cite{bittau2017prochlo} and ENPA by Apple and Google~\cite{apple2021exposure}. 
In Figure~\ref{fig:shuffle} (a) and (b), we show the differences between local and shuffle DP in their infrastructures.

\mypara{DP for network.}
Some works have been done to apply DP to protect users' privacy in network applications~\cite{sabzi2023netshaper,chang2022hide,zhang2019statistical,fan2021dpnet,liu2018epic}. 
For example, Sabzi et al. proposed NetShaper to buffer packets and make their traffic shape satisfy DP to defend against side-channel analysis~\cite{sabzi2023netshaper}. Chang et al. adds dummy DNS queries under local DP to defend against  DNS-based user tracking~\cite{chang2022hide}.
In this paper, we argue this direction of network-aware DP is also important and open questions exist.

\section{A Motivating Example}
\label{sec:example}

\begin{figure}
\centering
\includegraphics[width=.45\textwidth]{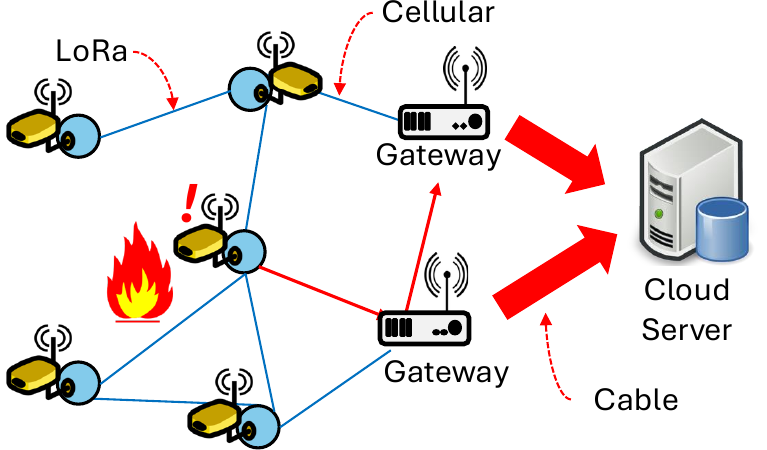}
\caption{Network of sensors for wildfire detection.
}
\label{fig:nodemesh}
\end{figure}

Here we describe a concrete setting that can benefit from DP and how the network setup can guide the design of DP mechanisms.
Imagine many camera censors are deployed in a rural area to monitor wildfire, as illustrated in Figure~\ref{fig:nodemesh}. 
The censors might accidentally capture the human activities of the monitored areas, which lead to linkage attack~\cite{di2013privacy}. To address such privacy threat, local DP or shuffle DP can be performed before the data are sent to the central cloud server.
Due to the limited network coverage in the rural area, a hybrid network structure is needed for data transmission. For instance, LoRa (Long Range) mesh network~\cite{de2017lorawan, lee2018monitoring} might be used by each sensor to relay data from its peers when the cellular network is unreliable, while normal gateways are still leveraged to transmit the data from the access networks to the core network. 
Such complex network setup introduces new research opportunities for DP and we will elaborate them as follows.

\section{DP under Untrusted Network}
\label{sec:security}

Since DP was born under the centralized setting (i.e., the central server is trusted) and the privacy is guaranteed with DP's post-processing property, the security analysis of DP mechanisms is relatively lacking.
Given the rise of decentralized DP, which assumes different trust models, some works investigated the attack surface under the theme of manipulation (or poisoning) attack~\cite{cheu2021manipulation, cao2021data, cheu2022differentially,li2023fine}, in which the attacker controls a fraction of corrupted users to send fake reports with extreme values and skew the estimation on the central server. 
While it is unsurprising that any data collection algorithms can be manipulated when users lie about their data, the prior works show that decentralized DP is more vulnerable when it is configured under high privacy requirement (i.e., small $\varepsilon$).
Yet, we found the network adversary was understudied, though it is expected to impact the the data transmission of any DP mechanism. Therefore, two research opportunities (ROs) naturally emerged.

\mypara{RO 1: Security analysis under the untrusted network.}
We identify two types of attacks that are relevant to network adversaries. First, the adversary could attempt to \textit{tamper} the network communications between users and the server to change the aggregated results. Under the setting of the motivating example (Section~\ref{sec:example}), the attacker would manipulate the packets from the sensors to force the server to mis-predict the wildfire locations. In this case, our threat model assumes standard network adversary, like attacker who controls a compromised access point, an insider within an ISP, a man-in-the-middle attacker in a wireless network, etc.

Different from the manipulation attacks using corrupted users~\cite{cheu2021manipulation, cao2021data, cheu2022differentially,li2023fine}, the network adversary has to work under more restrictions, e.g., that the packet content cannot be changed if users' reports are encrypted. Therefore, it would be interesting to assess whether/how the attacker can achieve the desired outcome with a limited set of capabilities like \textit{packet drop}, \textit{packet replay} and \textit{packet delay}, which do not break encryption mechanisms. 

The second type of adversary could sniff the network traffic and degrade the privacy guarantee of local DP or shuffle DP, and we assume the same threat model. In fact, a recent work showed that by analyzing the \textit{message timing, size and count} between the users and the shuffler, the probability of revealing the user-reported secret values can be increased by 90\%~\cite{wang2025side}. This attack is also applicable to the motivating example, such that the attacker can track the trajectory of a human victim from sensors' reports even when they are protected under DP. 

The combination of the two types of adversaries could lead to more powerful attacks, which has not been studied. We describe two scenarios as follows. First, fine-grained manipulation attack~\cite{li2023fine}, which aims to push the statistical result computed by the central to an intended value, would need to access the background information from the genuine users. This condition can be potentially met with traffic sniffing, so packet manipulation can be conducted strategically. Second, when DP is running on the infinite data streams~\cite{wang2021continuous}, the statistical analysis is time-sensitive (e.g., a range query that sums up the received values in a time period~\cite{wang2021continuous}). As such, the fine-grained manipulation attack in the stream setting would need to carefully select the time to manipulate the packets, which can be facilitated with traffic sniffing as well.

\mypara{RO 2: Network-centric defenses.}
Some mechanisms have been developed to defend decentralized DP. For instance, to defeat the manipulation attack, the server can actively detect the abnormal values reported by the users~\cite{li2023fine, murakami2025augmented} or directly detect the corrupted users~\cite{cao2021data}. The shuffler can perform dummy data addition ~\cite{wang2020improving, murakami2025augmented} and random sampling~\cite{murakami2025augmented} to mitigate the influence from the attacker. To defeat traffic sniffing, batching reports and message padding have been evaluated~\cite{wang2025side}. We believe the existing defenses need to be re-evaluated under the {\it new network adversaries}, and some defenses might be irrelevant. For example, when the attacker manipulates packets without corrupting users, detecting corrupted users~\cite{cao2021data} would be unnecessary. 

We believe adapting the network-centric defenses is a promising direction. For instance, adding report ID could handle the packet replay attack, and measuring the transmission timeout or latency could detect the adversarial packet drops and packet delays. That said, the defenses should be calibrated when the network environment is lossy to avoid false positives. For instance, packet losses can happen frequently under LoRa WAN, a network setup mentioned in the motivating example: Liu et al. conducted a study of packet losses in a standard LoRa network of Shanghai, China~\cite{liu2020characterizing}, and they found the packet loss rate (PLR) for most devices ranges from 0\% to 10\%, while it rises to \textit{over 90\%} for a few devices. Defenses tailored to network topologies could be investigated, and we describe a few ideas in Section~\ref{sec:topo}.

\begin{figure}
\centering
\includegraphics[width=0.45\textwidth]{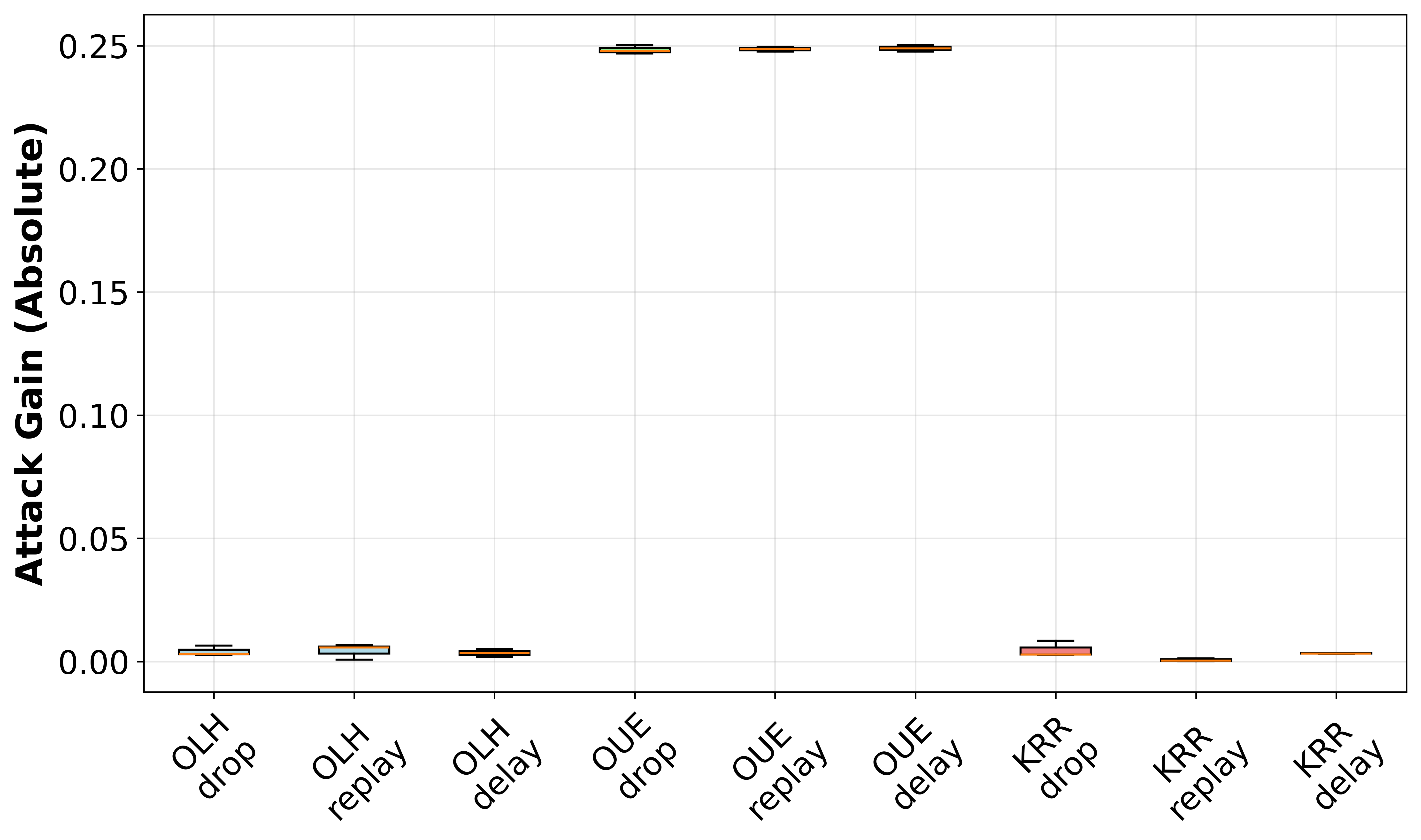}
\caption{Network RPA attack on the \texttt{uniform} dataset.
}
\label{fig:gain_rpa}
\end{figure}

\begin{figure}
\centering
\includegraphics[width=0.45\textwidth]{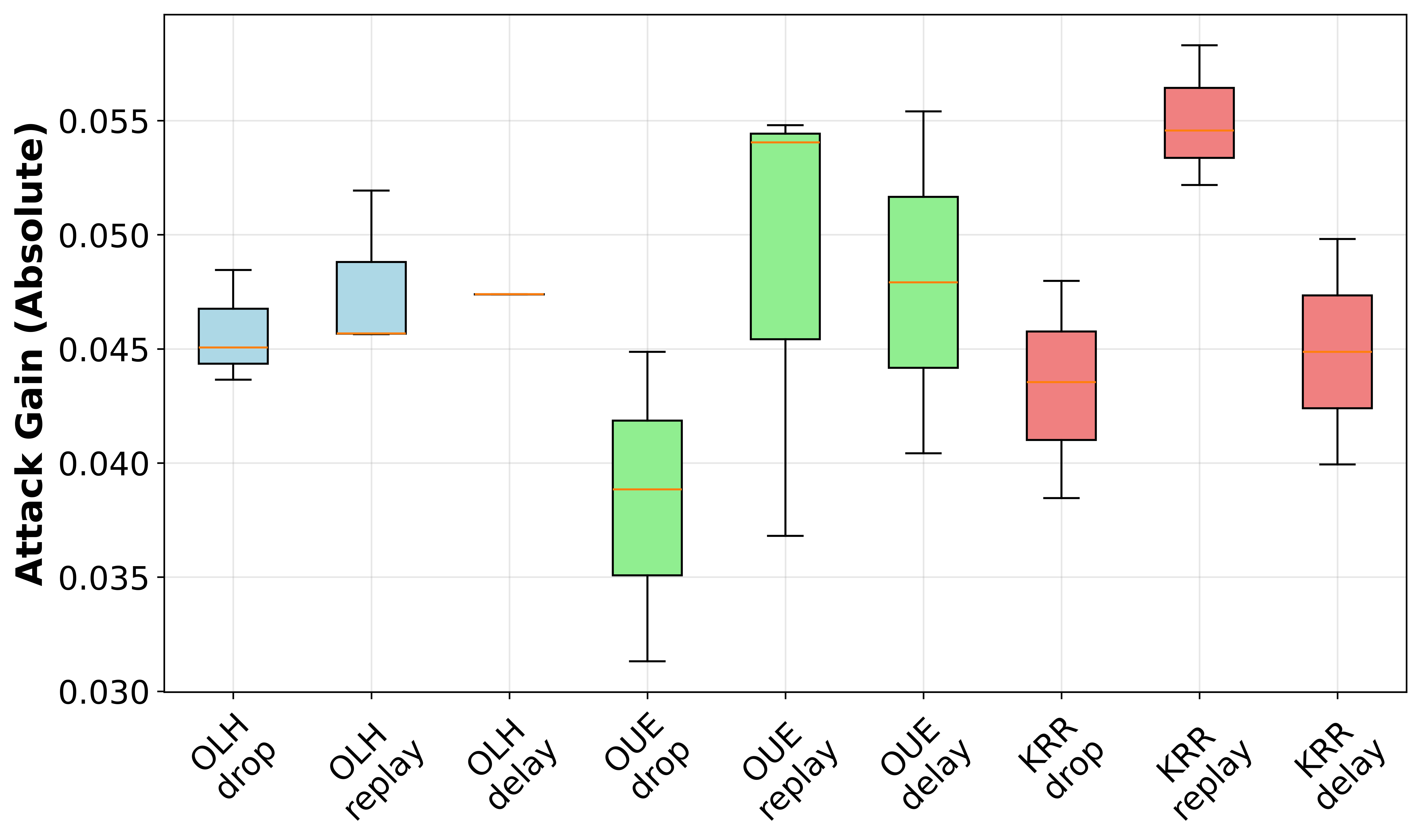}
\caption{Network RIA attack on the \texttt{uniform} dataset.
}
\label{fig:gain_ria}
\end{figure}

\mypara{Preliminary experiments.}
We performed a preliminary study about the feasibility of conducting network-based manipulation attacks against local DP protocols. We choose two attack methods proposed by Cao et al.~\cite{cao2021data}, namely random perturbed value attack (RPA), in which adversarial users report randomized perturbed values, and random item attack (RIA), in which adversarial users selectively promote target items. We extended the open-source implementation from LDPGuard~\cite{tkde/HuangOYHZZZZ24} to simulate packet drop, packet replay and packet delay on the clients' reports. Three local DP protocols for frequency estimation were evaluated, including optimized local hashing (OLH)~\cite{Wang2017usenix}, optimized unary encoding (OUE)~\cite{Wang2017usenix}, and $k$-ary randomized response (KRR)~\cite{kairouz2014extremal}.

Here we show the results on the \texttt{uniform} dataset synthesized by LDPGuard code~\cite{uniform_ldpguard}, which includes 128 items (represented by integers from 0 to 127) and 500,000 users, following the uniform distribution. The three attacks are configured as follows:
1) package drop: selective packet loss is simulated, parameterized by drop rates (0.1, 0.3 and 0.5), mimicking scenarios where an attacker can suppress user contributions;
2) package replay: previously observed packets are re-injected multiple times (5, 10, and 20), modeling adversarial inflation of statistical aggregates;
3) package delay: packets are delayed with varying ratios (0.2 and 0.5), affecting time-sensitive queries that aggregate user reports in a given period. 
Following~\cite{tkde/HuangOYHZZZZ24}, the attack effectiveness is mainly represented by \textit{attack gain}, which is the normalized difference between the estimation error under a local DP protocol before and after attack.

Figure~\ref{fig:gain_rpa} and ~\ref{fig:gain_ria} show the absolute values of attack gains of RPA and RIA on different combinations of network adversaries and local DP protocols. 
Similar attack effectiveness is observed as the baseline manipulation attack~\cite{tkde/HuangOYHZZZZ24}: 1) RIA reaches higher attack gain when kRR and OLH are attacked (also around 0.05 like ~\cite{tkde/HuangOYHZZZZ24}); 2) RPA is more effective when OUE is attacked (attack gain ranges between 0.2 to 0.5 in ~\cite{tkde/HuangOYHZZZZ24}), because OUE represents an input as a high-dimensional binary vector and RPA can perturb any bit without constraints. Meanwhile, we found packet replay is more effective than packet drop and packet delay, likely due to that packet replay can inject values to be aggregated by the central server. In short, the preliminary result shows network adversary should be considered under decentralized DP, and there is room to improve the attack techniques (e.g., finer-grained manipulation like~\cite{li2023fine} and stealthier manipulation on a smaller ratio of packets).

\section{Adapting DP to Network Topology}
\label{sec:topo}

\begin{figure*}[h]
    \centering
    \includegraphics[width=0.9\textwidth]{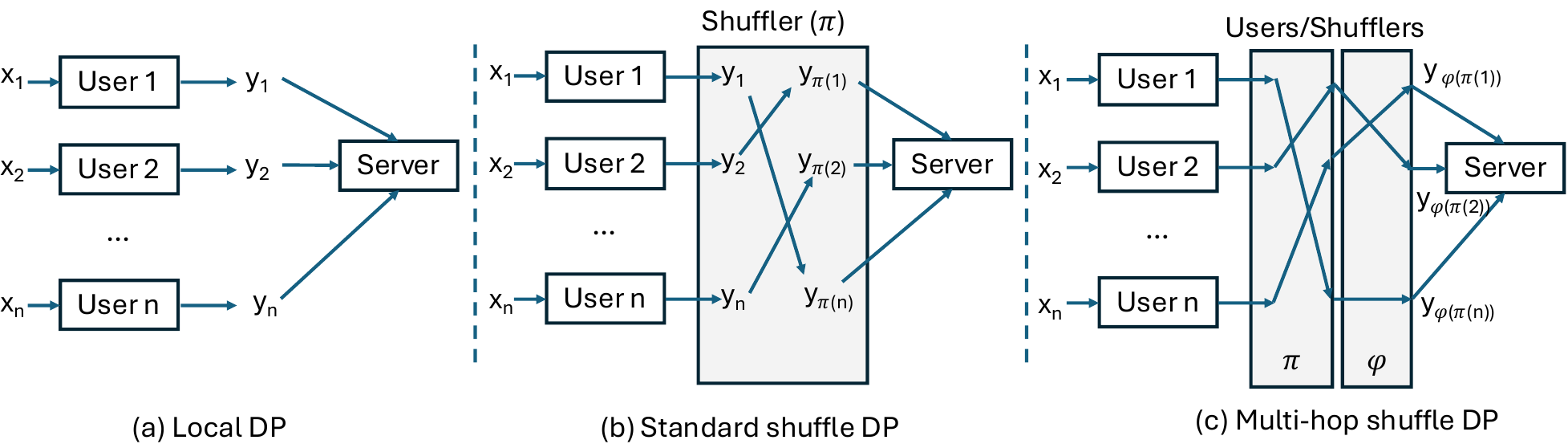}
    \caption{Comparison between local, standard shuffle and new multi-hop shuffle DP (described in RO3). For simplicity, we show 1-message shuffle DP~\cite{cheu2021differential}. $x_i$ and $y_i$ are the client's value before and after a local DP mechanism. $\pi$ and $\phi$ represent shufflers, which can also be users under multi-hop shuffle DP. 
    }
    \label{fig:shuffle}
\end{figure*}

Comparing to the central DP setting, decentralized DP involves a more complex network topology for data collection and aggregation. Yet, we found no prior work has systematically studied the impact of network topology on DP mechanisms, not to mention adapting DP mechanism to the underlying network topology. In this section, we focus on shuffle DP, as it requires a hierarchy of distant entities (i.e., clients, shuffler, and central server) to work together.

The first real-world deployment of shuffle DP is Prochlo by Google~\cite{bittau2017prochlo}, which assumes that all clients messages go to another shuffler before reaching the central server. This architecture has been followed by most of the shuffle DP works, but the shuffler creates a single point of failure. It also introduces high overhead to ensure message anonymity, due to the use of Trusted Executed Environment (TEE) such as Intel Software Guard Extension (SGX) to batch and shuffle reports from all users concurrently. 
Besides, TEE might not always be available in a data collection infrastructure. These constraints can be potentially resolved when the network topology is considered as a design factor. 
We describe the rationale below and elaborate three relevant ROs.

\mypara{RO 3: Message routing for shuffle DP.}
There are a few works investigating the alternative setups beyond centralized shuffler. For instance, concurrent shuffle lets the central server activate $k$ shufflers and the clients send messages to the $k$ shufflers concurrently in each round ~\cite{tenenbaum2023concurrent}. Network shuffling asks each user to exchange data in a random-walk fashion with the other users till reaching the central server ~\cite{liew2022network}. Comparing to Prochlo, network shuffling has much lower memory overhead ($O(1)$ vs $O(n)$ where $n$ is the number of users) at similar privacy amplification effects and moderate communication cost. 
Both network shuffling and concurrent shuffle address the issue of single point of failure by dynamically choosing a small set of shufflers from the entities between clients and central server (e.g., $O(\text{log}n)$ shufflers from $n$ clients by network shuffling).

Yet, none of the aforementioned works consider factors of a real-world network environment, and we believe a more principled framework is necessary to map the shuffle models to their deployed networks. Interestingly, these works all conduct some sorts of \textit{message routing} between clients and central server:  concurrent shuffle~\cite{tenenbaum2023concurrent} can be seen as performing \textit{multicast} routing, and random walk used by network shuffling~\cite{liew2022network} is a special case of \textit{multi-hop} routing, which is illustrated in Figure~\ref{fig:shuffle}. 
In fact, the motivating example can support the two schema by turning each sensor into a shuffler.  
Therefore, we draw the connections between network routing and shuffle DP and categorize the network-related design choices below.

\begin{itemize}
    \item \textbf{Types of addressing.} In addition to unicast (i.e., message sent to one centralized shuffler~\cite{bittau2017prochlo}) and multicast (i.e.,g $k$ concurrent shuffler~\cite{tenenbaum2023concurrent}), \textit{broadcast} and \textit{anycast} can be considered. Mapping anycast to shuffle DP can be particularly interesting as the selection of shuffler will be purely based on cost measures, which are elaborated later.
    
    \item \textbf{Routing algorithms.} When multiple shufflers are reachable from a client, the optimal sequence of shufflers can be determined under the cost attached to each shuffler and its policies (e.g., an edge device only shuffles the messages from the other devices manufactured by the same vendor). Routing the message can be more complex than random walk~\cite{liew2022network}, and the shuffler selection can be aided by the existing routing protocols like BGP and OSPF. 
    
    \item \textbf{Cost measures.} The cost of routing typically depends on the network-related metrics like bandwidth and latency. These metrics can be used to select a shuffler that fulfills certain requirements, e.g., low latency of message forwarding. The cost metrics could be further extended for privacy requirements. For example, as shuffler is supposed to ``de-link'' the client and its sent messages, we can measure its ``affinity'' to the client, e.g., by the number of shuffled messages per client, and deprioritize the high-affinity shufflers during shuffler selection.

    \item \textbf{Anonymous communication.} 
    The identities of the clients could be revealed under active adversaries who compromise shufflers or perform traffic analysis on messages from the clients to the central server. As a countermeasure, mix-net~\cite{chaum1981untraceable}, a classic anonymous communication framework was brought up to be used as the underlying primitive to realize shuffle DP~\cite{cheu2019distributed}. However, concerns were later raised~\cite{liew2022network} due to the efficiency drawback and the powerful attacks like flow correlation~\cite{attarian2023mixflow}. We argue that it is still worthwhile to investigate the more recent anonymous communication frameworks for shuffle DP, like Raceboat~\cite{vines2024communication}, which could bring better privacy guarantee and privacy amplification effects.

\end{itemize}

\mypara{RO 4: Shuffler discovery and verification.}
The previous shuffle DP works assume the clients and the central server have full knowledge about the usable shufflers (e.g., the central server selects $k$ shufflers in concurrent shuffle DP~\cite{tenenbaum2023concurrent}) and the shufflers are all reachable from the clients. However, these conditions might not always be satisfiable, e.g., when the client is moved to a region without reliable network connection to the Internet gateway. A potential solution is to find the nearby devices that are willing to relay and shuffle the messages for the client, and we call such devices ``open shufflers''. 
An interesting research question here is how to effectively discover the open shufflers and verify they fulfill the privacy guarantees desired by shuffle DP.

Regarding shuffler discovery, we expect a broad range of protocols (e.g., BLE mesh, multicast DNS, LLDP, and p2p DHT) can be retrofitted for this purpose. Take BLE Mesh as an example. A node volunteers to serve the role of open shuffler can broadcast a service advertisement indicating it can perform shuffling with other metadata like buffer capacity. The client nodes passively scan to gather the beacons, and select the open shufflers based on various metrics.

A more challenging issue is how to verify if an open shuffler behaves faithfully. Noticeably, this is less of a concern under Prochlo that builds the shuffler on TEE. We found an interesting connection between this issue with the research of verifying mixnet nodes, which has accumulated over a decade of works~\cite{haines2020sok}. Since the clients' messages are supposed to be encrypted (so message integrity is already protected), we can focus on verifying if messages are randomly permuted and relayed. Our initial investigation suggests three approaches summarized by ~\cite{haines2020sok} are relevant: 1) Randomized Partial Checking (RPC)~\cite{jakobsson2002making} which challenges a shuffler to reveal input-output links; 2) Trip wires~\cite{khazaei2012mix} that inject indistinguishable dummy messages to detect message dropping; 3) Zero-knowledge proof that asks a shuffler to prove its output is randomly permuted.

An orthogonal but important issue is how to incentivize device owner to run the open shuffler service. 
We found a recent work Nebula~\cite{watson2024nebula} pointed out an interesting solution, which incentivizes devices running Apple's FindMy network~\cite{findmy} or Amazon's Sidewalk~\cite{sidewalk} to backhaul sensor data to their registered servers with \textit{micro-payments}. We expect an ecosystem of open shufflers can be formed with the three aforementioned issues being addressed.

\mypara{RO 5: Network-aware message caching.}
Message caching is a well-established strategy to reduce the end-to-end communication latency, and we found caching has been examined in prior DP works, though for a different purpose. 
In the local DP setting, when a client caches a noisy answer locally and reuses it for the subsequent reports when the answer value is the same, more accurate estimation can be computed at the same privacy budget. 
In the central DP setting, by reusing past noisy responses to serve new queries that involve previous answers, privacy budget could be saved (hence answering more queries)~\cite{mazmudar2022cache}. 
So far, the cache implementations described by DP literature seem to be orthogonal to the caching mechanisms/infrastructure developed by the networking research, while we believe interesting designs can be embarked under the theme of network-aware DP caching. Below we describe a few ideas, by referencing DNS caches as an example.

\begin{itemize}
    \item{\textbf{Multi-tier caches.}} We can maintain caches both at the client side and the shuffler, just like local cache and resolver cache in DNS, to further reduce latency and improve privacy. 
    \item{\textbf{Caching algorithms and policies.}} The limited storage and data volatility are two main issues tackled by network caching research, and different caching algorithms (e.g. LRU and predictive pre-fetching) and policies (e.g., invalidating a cache entry after a Time-to-Live value). These two issues are also relevant to DP caches, which could be addressed by adapting the aforementioned solutions.
    \item{\textbf{Trust levels.}} Borrowing trust levels from DNS~\cite{bind9-trust}, we could differentiate source authenticity in DP caches, when multi-tier caches are deployed. For instance, when a client's reports are routed through multiple intermediate shufflers, DP caching protocols need to validate and prefer more reliable sources when inconsistencies occur, which resembles DNS in that authoritative answer if preferred over a cached one.
\end{itemize}

 \section{Conclusion and Discussion}
\label{sec:conclusion}

Through the lens of network research, we identified 5 research opportunities (ROs) that can improve decentralized DP and enable its deployment in complex network settings. We believe there is a large room for new works in the intersection between DP and network, and we hope to use this paper to attract more attention  to this interesting intersection. Below we discuss two relevant aspects not covered in the main body of the paper. 

\mypara{Protocol standardization.}
Given the increasing usage of decentralized DP for data collection by organizations in public and private sectors, we believe establishing a new standard or enhancing the existing standards for DP-powered data collection is necessary to mitigate implementation pitfalls. Distributed Aggregation Protocol (DAP)~\cite{geoghegan2023distributed} is the most relevant draft standard as far as we know, but it mainly focuses on validating clients' measurement in a privacy-preserving way. DP is briefly listed to protect the output privacy of the aggregated results, which resembles central DP. Hence, decentralized DP still needs standardization.

\mypara{Low-priority goals.}
One might notice that we did not discuss the ROs that reduce the bloated message size and count caused by DP mechanisms. Though theoretical analysis found some DP mechanisms have high asymptotic message complexities~\cite{wang2020comprehensive, cheu2021differential}, reducing the spatial overhead of messages might have limited real-world impact, due to that DP is often applied to the periodic reports sent by users and each report has a small size (e.g., emoji usage collected by Apple~\cite{apple-dp}). Yet, this direction can be revisited when DP is applied to large-volume and real-time data.

\section*{Acknowledgements}

We thank Dr. Thomas Steinke for the valuable suggestions. This work is supported by NSF CNS-2220434 and CNS-2220433. Any opinions, findings, and conclusions or recommendations expressed in this material are those of the author(s) and do not reflect the views of the sponsors.

\bibliographystyle{ACM-Reference-Format} 
\bibliography{main}

\end{document}